\documentclass[12pt,reqno]{amsart}
\usepackage{amsmath,amsfonts,amssymb,amscd,amsthm,amsbsy,epsf}
\usepackage[dvips]{graphicx}
\usepackage[usenames]{xcolor}
\usepackage{comment}
\pdfoutput=1 

\newtheorem{theorem}{Theorem}

\newtheorem{remark}[theorem]{Remark}

\newcommand\beq[1]{ \begin{equation}\label{#1} }
\newcommand{\eeq}{ \end{equation} }

\newcommand\beqa[1]{ \begin{eqnarray} \label{#1}}
\newcommand{\eeqa}{ \end{eqnarray} }
\newcommand{\beqano}{ \begin{eqnarray*} }
\newcommand{\eeqano}{ \end{eqnarray*} }
\newcommand\equ[1]{{\rm (\ref{#1})}}

\def\Ord{{\mathcal O}}

\def\integer{{\mathbb Z}}

\def\real{{\mathbb R}}
\def\torus{{\mathbb T}}

\begin{document}

\title[Hamiltonian formulation of the non-conservative spin-orbit model]{Hamiltonian formulation of
the spin-orbit model with time-varying non-conservative forces}

\author[I. Gkolias]{Ioannis Gkolias}
\address{
Department of Mathematics, University of Rome Tor Vergata, Via
della Ricerca Scientifica 1, 00133 Rome (Italy)}
\email{gkolias@mat.uniroma2.it}

\author[C. Efthymiopoulos]{Christos Efthymiopoulos}
\address{
Academy of Athens,Research Center of Astronomy and Applied Mathematics,Soranou Efessiou 4, GR-11527 Athens (Greece)}
\email{cefthim@academyofathens.gr}

\author[G. Pucacco]{Giuseppe Pucacco}
\address{
Department of Physics, University of Rome Tor Vergata, Via della Ricerca Scientifica 1,
00133 Rome (Italy)}
\email{pucacco@roma2.infn.it}

\author[A. Celletti]{Alessandra Celletti}
\address{
Department of Mathematics, University of Rome Tor Vergata, Via della Ricerca Scientifica 1,
00133 Rome (Italy)}
\email{celletti@mat.uniroma2.it}

\maketitle

\baselineskip=18pt              

\begin{abstract}
In a realistic scenario, the evolution of the rotational dynamics of a celestial or artificial body is subject to dissipative effects. Time-varying non-conservative forces can be due to, for example, a variation of the moments of inertia or to tidal interactions. In this work, we consider a simplified model describing the rotational dynamics, known as the \sl spin-orbit problem, \rm where we assume that the orbital motion is provided by a fixed Keplerian ellipse. We consider different examples in which a non-conservative force acts on the model and we propose an analytical method, which reduces the system to a Hamiltonian framework. In particular, we compute a time parametrisation in a series form, which allows us to transform the original system into a Hamiltonian one. We also provide applications of our method to study the rotational motion of a body with time-varying moments of inertia, e.g. an artificial satellite with flexible components, as well as subject to a tidal torque depending linearly on the velocity.
\end{abstract}

\maketitle

\vglue.1cm


\vglue.1cm

\section{Introduction}\label{sec:intro}

Celestial bodies usually rotate around a certain axis, which in many applications can be taken to nearly coincide with one of  the body's principal axes of inertia. In this case, the angular momentum $\Gamma$ is written as the product of the corresponding moment of inertia $C$ and the angular velocity $\dot{\theta}$, $\Gamma = C \dot{\theta}$. The time derivative of the angular momentum is
\begin{equation}
\frac{d \Gamma}{dt} = \frac{d C}{d t} \dot{\theta} + C
\ddot{\theta} = N_z, \label{eq:one}
\end{equation}
and thus its time evolution depends not only on the external torques $N_z$, but also on the time variation of the moment of inertia.

Since celestial bodies are not perfect rigid, variations in the moments of inertia can occur due to internal processes or from tidal deformations caused by the gravitational interactions with other bodies. For example, the gravitational attraction of the Moon and the Sun can tidally deform the Earth and cause variations in its polar moment of inertia $C$ (\cite{yoder1981}). A similar behaviour is observed also for artificial satellites; flexible components, like antennas and solar panels, for instance, can modify the shape or the mass distribution of the spacecraft and thus can alter the moments of inertia (\cite{inarrea2006}). 

The rotational dynamics can be studied by using a simplified approach known as the \sl spin--orbit model; \rm one considers a small body, e.g., a satellite, moving around a primary body on a fixed Keplerian ellipse and rotating around an internal axis, which is assumed to be perpendicular to the orbit plane. Moreover, one takes the spin-axis as coinciding with the smallest physical axis. Thus, denoting by $A<B<C$ the principal moments of inertia, the spin-axis corresponds to $C$. The time variations in the moment of inertia $C(t)$ introduce an angular velocity-dependent term in the equations of motion for the body's rotation\footnote{In general the other two axes of inertia, $A$ and $B$, might also vary in time, say $A=A(t)$, $B=B(t)$, but in the present setting this variation does not contribute to Eq.~(\ref{eq:one}), where only the variation of $C$ is considered.}. 

Dissipative terms in general destroy the symplectic structure of the equation of motion describing the spin-orbit model; the symplectic structure would hold in the presence of solely external gravitational torques in a rigid-body configuration. On the other hand, a symplectic formulation of the equation of motion might be a desirable feature in many theoretical or numerical studies of spin-orbit models. An example was provided by Henrard (\cite{henrard1993}), who proposed a suitable time parametrisation which, in the case of a constant drag force, can map the dissipative system into a Hamiltonian one, through a non-canonical transformation. This idea became the cornerstone of the adiabatic invariant theory that explains the trapping into the resonances of the spin-orbit evolution (\cite{yoder1979}, \cite{henrard1982}, see also \cite{CCbasin}). In addition, this mapping allows us to use the whole suite of formal tools from Hamiltonian theory to tackle a system of manifestly non-Hamiltonian character in its original formulation.

It is straightforward to see that the idea of the time parametrisation can also be applied in the case of a time varying dissipative term, as described in Sec. \ref{sec:theory}, thus accounting for variations of the moment of inertia $C$. Given the functional form of these variations, the problem then reduces to a second-order differential equation, the solution of which gives the suitable time parametrisation to map the system into a Hamiltonian one.

In the present paper our aim is to develop an algorithmic procedure, based on a series approach, that gives us the time parametrisation in an explicit form in cases more generic than the one considered by Henrard in \cite{henrard1993}. In particular, we extend the computation in the cases of a quasi-periodic variation of $C$, as well as the more general case of a dissipative term with both a constant and a quasi-periodic contribution. 

We apply our theory to the study of three different applications: i) the rotational motion of an artificial satellite with flexible components, when the periodic variation in $C$ is resonant with the orbital motion, or ii) when it is non-resonant with the orbital motion, and iii) the rotational motion of a celestial body under the effect of a constant (tidal) and a quasi-periodic dissipation. These models can be conveniently described by a second order differential equation of the form $\ddot{\theta} = G(\theta,t) + F(t)\dot{\theta}$  (see Eq.~(\ref{eq:orsystem}) below), for some analytical functions $G$ and $F$. 

Two main advantages of the method presented here are: i) the so-computed series lead to an analytical representation of the solution whose accuracy nearly reaches machine precision after only a few iterative steps. ii) The same representation allows, via the Hamiltonian re-formulation, to use symplectic integrator techniques in order to compute numerically the dynamical evolution of systems whose original formulation is non-Hamiltonian. This, in turn, allows to take benefit of all the advantages associated with the use of symplectic integration techniques, as exposed, e.g., in \cite{Galley}, \cite{Tsang}, \cite{mac}. Of course, the trade-off is the need to make symbolic series computations, whose realization, however, exhibits no difficulty with the use of a standard symbolic manipulator.

This paper is organised as follows: in Sec.~\ref{sec:theory} we present the theory for the computation of a suitable time parametrisation used to cast a particular class of dissipative systems into a Hamiltonian one; in Sec.~\ref{sec:applications} we discuss some interesting applications of our method and in Sec.~\ref{sec:conclusions} we provide our conclusions.

\section{Theory of time parametrisation}\label{sec:theory}

In this section we consider a dynamical system described by the equation

\begin{equation}
\ddot{\theta} = G(\theta,t) + F(t) \dot{\theta}\ ,
\label{eq:orsystem}
\end{equation}
where $\theta \in \torus$, $t \in \real$, $G$, $F$ are analytic
functions with $G$ periodic; moreover, $G$ is assumed to be minus the gradient of a potential function $V(\theta,t)$:
$$
G(\theta,t) = - \frac{\partial V(\theta,t)}{\partial \theta}.
$$
Equation~(\ref{eq:orsystem}) describes, for example, the rotational motion of a satellite under suitable assumptions, the so-called ``spin-orbit" model (see, e.g., \cite{celletti1990}, \cite{celletti1990b}) and subject to a tidal torque, depending linearly on the velocity or rather admitting a flexible structure, so that the moment of inertia $C$ varies with time. We describe how to obtain a time parametrisation, such that the equations of motion can be associated to a Hamiltonian function.

Let us write Eq. (\ref{eq:orsystem}) as the first order differential system:

\begin{align}
\dot{\theta} &= p, \nonumber \\
\dot{p} &= G(\theta,t)+ F(t)\  p,
\label{eq:originalsystem}
\end{align}
where $p, t \in \real$, $\theta\in\torus$.

Let us introduce a time parametrisation
$$
\tau=\tau(t),
$$
where the function $\tau=\tau(t)$, or equivalently $t=t(\tau)$, will be supposed to satisfy a suitable differential equation (see Eq. (\ref{eq:deqtimepar}) below). We introduce a new generalised momentum as
$$
I=\frac{d\theta}{d\tau}.
$$
The equations of motion in the new time $\tau$ read
\begin{equation}
I=\frac{d\theta}{d\tau} = \frac{d\theta}{dt} \frac{dt}{d\tau} = \dot{\theta} \frac{dt}{d\tau}.
\label{eq:neweq1}
\end{equation}
From (\ref{eq:neweq1}) we obtain that the variation of $I$ in terms of the time $\tau$ is given by
\begin{equation}
\frac{dI}{d\tau}=\frac{d^2\theta}{d\tau^2} = \frac{d}{d\tau} \left(  \frac{d\theta}{dt} \frac{dt}{d\tau} \right)  = \frac{d\theta}{dt} \frac{d^2 t}{d\tau^2}+\frac{d^2 \theta}{dt^2} \left(\frac{dt}{d\tau} \right)^2.
\label{eq:neweq2}
\end{equation}
Substituting Eq.~(\ref{eq:originalsystem}) into Eq.~(\ref{eq:neweq2}) yields
$$
\frac{dI}{d\tau} = \dot{\theta}  \frac{d^2 t}{d\tau^2} + \left( G(\theta,t)+ F(t) \dot{\theta} \right) \left(\frac{dt}{d\tau} \right)^2.
$$
A rearrangement of the terms gives:
$$
\frac{dI}{d\tau} =  G(\theta,t)  \left(\frac{dt}{d\tau} \right)^2 + \dot{\theta} \left( \frac{d^2 t}{d\tau^2}+F(t) \left(\frac{dt}{d\tau} \right)^2 \right).
$$
If we define a time parametrisation such that:
\begin{equation}
\frac{d^2 t}{d\tau^2}+ F(t) \left(\frac{dt}{d\tau} \right)^2 = 0,
\label{eq:deqtimepar}
\end{equation}
the equations of motion become:
\begin{align}
\frac{d \theta}{d \tau} &= I, \nonumber \\
\frac{dI}{d\tau} & =  G(\theta,t(\tau))  \left(\frac{dt(\tau)}{d\tau} \right)^2,
\label{eq:equationsham}
\end{align}
where now $t = t(\tau)$, i.e., the time $t$ is expressed as a function of the new time $\tau$ through the solution of Eq.~(\ref{eq:deqtimepar}). Equations (\ref{eq:equationsham}) are Hamilton's equations under the Hamiltonian function
\begin{equation}
H(I,\theta,\tau) = \frac{I^2}{2} + V(\theta,t(\tau))  \left(\frac{dt}{d\tau} \right)^2.
\label{eq:ham}
\end{equation}
Thus, given a suitable solution of the differential Eq.~(\ref{eq:deqtimepar}), we can time-parametrise the equations of motion and bring the system to Hamiltonian form. The advantage of this approach is that we can now apply a variety of tools developed in the framework of the  Hamiltonian theory to formally study the system.

The fact that a dissipative system is mapped into a Hamiltonian one, via the time parametrisation satisfying Eq.~(\ref{eq:deqtimepar}), may at first seem counter-intuitive. A simple consideration to explain this idea was offered in \cite{henrard1993}. One notes that the new momentum $I$ is time dependent (Eq.~(\ref{eq:neweq1})). Thus, the conservation of area in the phase space ($\theta,I$) corresponds to a time dependent evolution of the area of the phase space ($\theta,p$), which, in turn, depends on the form of $F(t)$. The idea of exploiting the time-dependent mapping of areas in phase space was already discussed in the pioneer work of Andronov et. al. \cite{Andronov1966}.

In the following we will demonstrate how the analytical solution of Eq.~(\ref{eq:deqtimepar}) is obtained in three different scenarios: (I) in the case where $F$ is constant (Sec.~\ref{subsec:theory1}), (II) in the case where $F$ is a quasi-periodic function of time (Sec.~\ref{subsec:theory2}) and (III) in the more general case where the function $F$ is the sum of constant and quasi-periodic terms (Sec.~\ref{subsec:fulltheory}). Case (I) is the one treated in \cite{henrard1993}, while cases (II) and (III) will be treated by a new approach based on series expansions.

\subsection{I. $F$ constant}\label{subsec:theory1}

Let us assume that $F$ has a constant value $F(t)=a$. This is exactly the problem treated by Henrard (see \cite{henrard1993} and references therein). Eq.~(\ref{eq:deqtimepar}) takes the form:
$$
\frac{d^2 t}{d\tau^2} + a \left(\frac{dt}{d\tau} \right)^2 = 0,
$$
and the general solution is 
\begin{equation}
t = C_1 + \frac{ \log( a \tau + C_2)}{a} ,
\label{eq:gentoftaucase1}
\end{equation}
where $C_1,C_2$ are real integration constants. Eq.~(\ref{eq:gentoftaucase1}) defines a two-parametric family of possible solutions that satisfy Eq.~(\ref{eq:deqtimepar}), however we only need one of these solutions. We will see in the following paragraphs that the selection of the constants plays a crucial role in constructing a reliable scheme. In this simplified example, one can just choose $C_1=0,C_2=1$ leading to
$$
t = \frac{ \log(a \tau + 1 ) }{a},
$$
or
$$
\tau = \frac{e^{at}-1}{a}.
$$
With this choice one has that $\tau=0$ at $t=0$. The Hamiltonian function that models our system in this case takes the form:

\begin{equation}
H(I,\theta,\tau) = \frac{I^2}{2} + (a \tau+1)^{-2} V(\theta,t(\tau)). \label{eq:ham2}
\end{equation}
Notice that in the above formulation the quantity
$$
\left(\frac{dt}{d\tau} \right)^2=(a \tau+1)^{-2},
$$
is slowly varying, since $a$, the dissipation constant, is usually a small parameter in astronomical applications. Eq.~(\ref{eq:ham2}) has been used to model a wide variety of physical systems. For example, it is the core of the adiabatic invariant theory (\cite{henrard1993},\cite{batygin2015}), that has been used to explain the trapping of bodies into a spin-orbit resonance, namely a commensurability between the period of rotation of the satellite and the period of revolution around the main body.

\subsection{II. $F$ quasi-periodic }\label{subsec:theory2}

Let us assume that $F(t)$ has a quasi-periodic form admitting a decomposition in trigonometric series, say
\begin{equation}
F(t) = \sum_k (b_k \cos{\omega_k t} +c_k \sin{\omega_k t} ),
\label{eq:fourdecomp}
\end{equation}
where $b_k,c_k$ are small coefficients and $\omega_k$ are the corresponding frequencies. The frequencies $\omega_k$ may be either incommensurable or satisfy one or more commensurability conditions of the form 
$$
\sum_k m_k \omega_k = 0,
$$
with $m_k$ integers. The number of frequencies can, in principle, be arbitrary, while, as shown below, our method works equally well independently of the number of commensurabilities among the frequencies $\omega_k$. 

Similarly to the constant $a$ in case (I), we will assume that the amplitudes $b_k,c_k$ in Eq.~(\ref{eq:fourdecomp}) are small quantities. Then, our goal will be to find a solution of Eq.~(\ref{eq:deqtimepar}) using a series method. To this end, introducing the variable
\begin{equation}
v = \frac{dt}{d\tau},
\label{eq:v}
\end{equation}
Eq.~(\ref{eq:deqtimepar}) takes the form
\begin{equation}
\frac{dv}{d\tau} + F(t(\tau)) v^2 = 0.
\label{eq:deqtime2}
\end{equation}
Applying the chain rule
$$
\frac{dv}{d\tau} = \frac{dv}{dt} \frac{dt}{d\tau} = \frac{dv}{dt} v,
$$
Eq.~(\ref{eq:deqtime2}) yields
$$
\frac{dv}{dt} + F(t) v = 0,
$$
and thus the general solution for $v$ is
$$
v = e^{-(\int F(t) dt + C_1)},
$$
where $C_1$ is an integration constant. We substitute $v$ back from Eq.~(\ref{eq:v}) to get
$$
\frac{dt}{d\tau} = e^{- (\int F(t) dt + C_1)},
$$
and after one more integration we have
\begin{equation}
\tau + C_2 = \int e^{\int F(t)dt + C_1} dt = \int e^{ w(t) + C_1} dt,
\label{eq:solution}
\end{equation}
where $w(t) = \int F(t)dt$ and $C_2$ is a second integration constant.

Equation (\ref{eq:solution}) allows, now, to define a reliable scheme to calculate the time parametrisation for a quasi-periodic function $F(t)$ of the form (\ref{eq:fourdecomp}). The  algorithmic process can be divided in the five steps described below.

\subsubsection{Step 1} We first introduce a 'book-keeping' parameter $\lambda$ (see \cite{efthymiopoulos2011}), whose presence allows to organize the whole computation in powers of the quantity $\lambda$. Thus we rewrite Eq.~(\ref{eq:fourdecomp}) as
$$
F(t) = \lambda \sum_k (b_k \cos{\omega_k t} +c_k \sin{\omega_k t} ),
$$
where the coefficients $b_k,c_k$ are rescaled so as to have values of order unity, while $\lambda$ is now the small parameter. Then, in the first step we compute $w(t)$ as
\begin{align}
\label{eq:wt} w(t) &= \int F(t) dt = \lambda \int \sum_k ( b_k
\cos{\omega_k t} + c_k \sin{\omega_k t} ) dt  \\ \nonumber &=
\lambda \sum_k ( M_k \cos{\omega_k t} + N_k \sin{\omega_k t}),
\end{align}
where $M_k,N_k$ are suitable constants.

\subsubsection{Step 2} In the next step we compute the exponential of $w(t)$ through a Taylor expansion up to some maximum truncation order
$n$. Precisely, we have:
$$
e^{w(t)} = 1 + w(t) +  \frac{1}{2} w^2(t) + \frac{1}{3!} w^3(t) + \ldots + \frac{1}{n!} w^n(t) + \Ord(w^{n+1}),
$$
which leads to
\begin{equation}
e^{w(t)} = 1 + \sum_{0<s\leq n} z_s \lambda^s + \sum_{0<s \leq n}
\lambda^{s} \sum_k (E_k \cos{\Omega_k t} + F_k \sin{\Omega_k t})+\Ord(\lambda^{n+1}),
\label{eq:expwt}
\end{equation}
where $z_s$, $E_k$, $F_k$ are real constants. Note that the frequencies $\Omega_k$ of (\ref{eq:expwt}) come from linear combinations of the form $\sum_k m_k \omega_k$ for $m_k$ integers and with $\sum |m_k| \leq n$, where $\omega_{k}$ are any of the frequencies appearing in (\ref{eq:wt}). However the most important remark is that, in addition to the trigonometric terms, some constant terms also appear in (\ref{eq:expwt}). These constant terms are problematic, because they will produce secular terms in our final solution. Therefore we need to find a way to eliminate them. This is possible by adjusting the value of the constant $C_1$, as we are going to show in Step 3 below.

\subsubsection{Step 3}

In order to eliminate the terms $\sum_{0<s \leq n} z_s \lambda^s$ we set
$$
e^{C_1} = \frac{1}{1+ \sum_{0<s \leq n} z_s \lambda^s  +\Ord(\lambda^{n+1})},
$$
so that one has
\begin{equation}
e^{w(t)+C_1} = \frac{ 1 + \sum_{0<s \leq n} z_s \lambda^s +
\sum_{0<s \leq n} \lambda^{s} \sum_k (E_k \cos{\Omega_k t} + F_k
\sin{\Omega_k t}) +\Ord(\lambda^{n+1})}{1+ \sum_{0<s \leq n} z_s \lambda^s  +\Ord(\lambda^{n+1})}.
\label{eq:eliminatesec}
\end{equation}
Note that, for $\lambda$ sufficiently small, the denominator in (\ref{eq:eliminatesec}) is safely bound away from zero. Expanding Eq.~(\ref{eq:eliminatesec}) in powers of $\lambda$ up to order $n$ yields
\begin{equation}
e^{w(t)+C_1} = 1 + \sum_{0<s \leq n} \lambda^s \sum_k (G_k
\cos{\Omega_k t} + H_k \sin{\Omega_k t})+\Ord(\lambda^{n+1}),
\label{eq:expwtc1}
\end{equation}
for suitable constants $G_k,H_k$. Eq.~(\ref{eq:expwtc1}) is now an expression free of secular terms. Furthermore, all the coefficients and frequencies appearing in Eqs.~(\ref{eq:expwt}) to (\ref{eq:expwtc1}) are explicitly computable in terms of the original coefficients, $b_k$ , $c_k$ and the frequencies $\omega_k$ appearing in the definition of $F(t)$ given by Eq.~(\ref{eq:fourdecomp}).

\subsubsection{Step 4}

In this step we compute the solution $\tau(t)$ in
Eq.~(\ref{eq:solution}) as
$$
\tau + C_2 = \int e^{w(t)+C_1} dt = \int \left( 1 + \sum_{0<s \leq
n} \lambda^s \sum_k (G_k \cos{\Omega_k t} + H_k \sin{\Omega_k
t})\right) dt+\Ord(\lambda^{n+1}),
$$
which provides
\begin{equation}
\tau + C_2 = t + \sum_{0<s \leq n} \lambda^s \sum_k (P_k
\cos{\Omega_k t} + Q_k \sin{\Omega_k t})+\Ord(\lambda^{n+1}). \label{eq:tauoftcase2}
\end{equation}
The value of $C_2$ is computed from the requirement that at $t=0$, $\tau$ be also equal to zero. Note that similarly to $e^{C_1}$, $C_2$ also appears as a series in powers of the parameter $\lambda$.

\subsubsection{Step 5} As a final step we need to determine $t(\tau)$ in order to implement the time parametrisation.
We compute the series $t(\tau)$ by inverting the series of $\tau(t)$. In order to accomplish this task, first we rewrite Eq.~(\ref{eq:tauoftcase2}) as

\begin{equation}
t = \tau + C_2 - \sum_{0<s \leq n} \lambda^s \sum_k (P_k
\cos{\Omega_k t} + Q_k \sin{\Omega_k t})+\Ord(\lambda^{n+1}).
\label{eq:transformcase2}
\end{equation}

Then we substitute Eq.~(\ref{eq:transformcase2}) into itself $n$ times and we expand in Taylor series up to order $n$ in $\lambda$. This results in the following final expression for $t(\tau)$:
\begin{equation}
t = \tau +  C'_2 + \sum_{0<s \leq n} \lambda^s \sum_k (R_k
\cos{v_k \tau} + S_k \sin{v_k \tau})+\Ord(\lambda^{n+1}),
 \label{eq:soltcase2}
\end{equation}
where $C_2'$ is a real constant and where for suitable constants $R_k,S_k$, anew, the frequencies $v_k$ stem from linear combinations of the frequencies $\Omega_k$ with integer coefficients.

\begin{remark}
Steps 1-5 are formal. We have no rigorous proof of convergence. However, many examples (compare with Sec.~\ref{sec:applications}) show a very good convergence behaviour at least up to a truncation order suitable for all practical purposes.
\end{remark}

\begin{remark}
In Step 5, we use the back-substitution method to invert the series. There exist several more sophisticated methods for series inversion (see e.g. Press et. al. \cite{Press1982}). However, such methods are designed, in general, for either purely polynomial or purely trigonometric series, while the series presently dealt with simultaneously contain, in general, terms linear, trigonometric, and even exponential in t (see Sec.~\ref{subsec:fulltheory}). This implies that their inversion results in series which contain terms linear, trigonometric and logarithmic in the new time $\tau$. Thus, such series seem possible to invert only via the back-substitution algorithm.
\end{remark}

\subsection{III. $F$ with both a constant and quasi-periodic component}\label{subsec:fulltheory}
In this section we will treat the more general case where the function $F(t)$ has both a constant and a quasi-periodic component

\begin{equation}
F(t) = a +  \lambda \sum_k ( b_k \cos{\omega_k t} + c_k
\sin{\omega_k t} ). \label{eq:F}
\end{equation}

Following the discussion of the previous subsection, the time parametrisation is given from the equation
\begin{equation}
\tau + C_2 = \int e^{\int F(t)dt + C_1} dt  = \int e^{at + w(t) + C_1} dt,
\label{eq:solution2}
\end{equation}
where, taking into account \equ{eq:F}, we have:
$$
w(t) = \lambda \int \sum_k ( b_k \cos{\omega_k t} + c_k \sin{\omega_k t} ) dt.
$$
As we are going to show below, the algorithmic process to derive the time parametrisation is similar to the one described in the previous section.

\subsubsection{Steps 1-3}

The first three steps are identical to those of Sec.~\ref{subsec:theory2}, therefore their application leads to
\begin{equation}
e^{w(t)+C_1} = 1 + \sum_{0<s \leq n} \lambda^s \sum_k (G_k
\cos{\Omega_k t} + H_k \sin{\Omega_k t})  +\Ord(\lambda^{n+1}), \label{eq:solution3new}
\end{equation}
for suitable real constants $G_k$, $H_k$ and for $\Omega_k$ as in Step 2 of Sec.~\ref{subsec:theory2}.

\subsubsection{Step 4}

Substituting Eq.~(\ref{eq:solution3new}) into Eq.~(\ref{eq:solution2}), we have
$$
\tau + C_2 = \int e^{at + w(t) + C_1} dt = \int e^{at} \left(  1 +
\sum_{0<s \leq n} \lambda^s \sum_k (G_k \cos{\Omega_k t} + H_k
\sin{\Omega_k t}) \right) dt  +\Ord(\lambda^{n+1}),
$$
which yields
\begin{equation}
\tau + C_2 = e^{at} \left( \frac{1}{a} +  \sum_{0<s \leq n}
\lambda^s \sum_k (P_k \cos{\Omega_k t} + Q_k \sin{\Omega_k t})
\right) +\Ord(\lambda^{n+1}), \label{eq:tauall}
\end{equation}
where $P_k,Q_k$ are suitable constants.
\subsubsection{Step 5}

Once obtained $\tau(t)$ through Eq.~(\ref{eq:tauall}), we need an inversion to get $t(\tau)$. For this reason we rewrite Eq.~(\ref{eq:tauall}) as

\begin{equation}
\frac{1}{a} \log{\left(a\tau + C'_2\right)} = t + \frac{1}{a}
\log{\left(1 + \sum_{0<s \leq n} \lambda^s \sum_k (P_k
\cos{\Omega_k t} + Q_k \sin{\Omega_k t}) \right)} +\Ord(\lambda^{n+1}), \label{eq:taun}
\end{equation}
where $C_2'=a\, C_2$.

Then, we expand the logarithm of the trigonometric series once again in Taylor series of $\lambda$ up to order $n$, so to obtain
\begin{align}
\frac{1}{a} \log{\left(1 + \sum_{0<s \leq n} \lambda^s \sum_k (P_k \cos{\Omega_k t} + Q_k \sin{\Omega_k t})\right)} = \nonumber\\
= \sum_{0<s \leq n} \lambda^s \sum_k (R_k \cos{\Omega_k t} + S_k
\sin{\Omega_k t}), \label{eq:taun2}
\end{align}
for suitable coefficients $R_k$, $S_k$. From Eq.s (\ref{eq:taun}) and (\ref{eq:taun2}), we obtain
\begin{equation}
t = \frac{1}{a} \log{\left(a\tau + C'_2\right)} - \sum_{0<s \leq
n} \lambda^s \sum_k (R_k \cos{\Omega_k t} + S_k \sin{\Omega_k t}) +\Ord(\lambda^{n+1}).
\label{eq:tt}
\end{equation}

Substituting Eq.~(\ref{eq:tt}) into itself $n$ times and expanding in powers of $\lambda$ up to order $n$, we can invert the series and obtain finally $t$ as a function of $\tau$. We note the peculiarity of this expression, with respect to the corresponding one (Eq.~\ref{eq:soltcase2}) in the absence of dissipation. Namely, as shown in the example of Sec.~\ref{sec:tidal}, in the series function representing
$t(\tau)$ there appear trigonometric terms with argument
$$
L=\frac{\log\left( 1+a \tau \right)}{a},
$$
(compare with \equ{eq:step53} below).
Then, the terms in the new series have the general form:
$$
\frac{\Pi_1(a,\tau) \lambda^{k_1} \cos^{k_2}{L} \sin^{k_3}{L}}{\Pi_2(a,\tau)},
$$
where $\Pi_1,\Pi_2$ are polynomials in $a,\tau$ and $k_1,k_2,k_3$ integers.

Again, in this formal scheme, we do not have a rigorous proof of convergence. Nevertheless, we found numerical indications of a good convergence behaviour in all studied numerical examples.

\section{Applications}\label{sec:applications}

In this section we will discuss three different examples, in which the analytical theory developed in Sec.~\ref{sec:theory} is applied. We recall that the rotation of a body around its principal axis of inertia $C$ is described by Euler's equation

\begin{equation}
\frac{d \Gamma}{d t} =  \frac{d C}{d t} \dot{\theta} + C \ddot{\theta} = N_z,
\label{eq:euler}
\end{equation}
where $\Gamma$ is the angular momentum, $\dot{\theta}$ is the angular velocity and $N_z$ is the sum of the external torques. The gravitational torque for a triaxial rigid body orbiting on a Keplerian ellipse around a point mass perturber is given by the equation (\cite{murray1999}):
\begin{equation}
N_{z(triaxial)} =- \frac{3}{2} (B - A) \nu \left(\frac{\alpha}{r} \right)^3 \sin{(2 \theta - 2 f )},
\label{eq:nztriaxial}
\end{equation}
where $A<B<C$ are the moments of inertia in the body fixed frame, $\alpha$ is the semi-major axis, $\nu$ is the orbital frequency, $r$ is the distance of the two bodies and $f$ is the true anomaly. The functions $r(t)$, $f(t)$ are periodic functions of time and we can Fourier-decompose them, thus transforming (\ref{eq:nztriaxial}) into
$$
N_{z(triaxial)}(\theta,t) = - \frac{3}{2} (B - A) \nu \sum_{m \neq
0,m=-\infty}^{m=\infty} W\left(\frac{m}{2},e\right) \sin(2 \theta-
m t ),
$$
where the coefficients $W=W\left(\frac{m}{2},e\right)$ are series in the eccentricity known as Cayley coefficients (\cite{cayley1861}).  In addition to the gravitational torque, we will now consider three different examples of an additional torque. In the first two examples, we will consider only the angular velocity-dependent torque due to the variation in the moment of inertia $C$. In the first example we let $C(t)$ be periodic in resonance with the orbital motion, while in the second example we study the non-resonant case. In the third example we consider the evolution under both the torque from the time varying moment of inertia $C$ and a torque due to the third body perturbation of the form
\begin{equation}
N_{z(tidal)}(\dot\theta) = \mu + a \dot{\theta},
\end{equation}
with $\mu$, $a$ real constants and $\mu>0$, $a<0$, as motivated by the models of tidal torques in \cite{goldreich1966}, \cite{efroimsky2009}, \cite{noyelles2014}.

\subsection{Example of $C(t)$ periodic in resonance with the orbit.}

First, we consider a body (e.g., an artificial satellite) with a moment of inertia $C$ which varies periodically in time around an average value $\hat{C}$ with frequency $\Omega$:
\begin{equation}
C(t) = \hat{C} + \lambda \cos( \Omega t).
\label{eq:coft}
\end{equation}
A model of the type (\ref{eq:coft}) has been used in the past (see \cite{inarrea2006} and references therein) to describe a spacecraft with flexible components. A similar approach is also used in \cite{noyelles2016}, where the elements of the tensor of inertia are expressed as the sum of a frozen (constant) and a hydrostatic (time-varying) component. The differential equation describing the rotation is
$$
\ddot{\theta} = \frac{N_z}{C} - \frac{dC}{dt} \frac{\dot{\theta}}{C},
$$
or
\begin{equation}
\ddot{\theta} =  - \frac{3}{2} \frac{\nu (B - A)}{\hat{C} +
\lambda \cos (\Omega t)}  \sum_{m \neq 0,m=-\infty}^{m=\infty}
W\left(\frac{m}{2},e\right) \sin(2 \theta- m t ) + \frac{\lambda
\Omega \sin{(\Omega t)}}{\hat{C} + \lambda \cos(\Omega
t)}\dot{\theta}.
\label{eq:model1}
\end{equation}

The above equation is equivalent to Eq.~(\ref{eq:originalsystem}) with
\begin{align}
G(\theta,t) &=  - \frac{3}{2} \frac{\nu (B - A)}{\hat{C} + \lambda \cos (\Omega t)}  \sum_{m \neq 0,m=-\infty}^{m=\infty} W\left(\frac{m}{2},e\right) \sin(2 \theta- m t ),\nonumber\\
F(t) &= \frac{\lambda \Omega \sin{(\Omega t)}}{\hat{C} + \lambda
\cos(\Omega t)}.\nonumber
\end{align}
The potential function $V(\theta,t)$ associated with $G(\theta,t)$ is
$$
V(\theta,t) = - \frac{3}{4} \frac{\nu (B - A)}{\hat{C} + \lambda \cos (\Omega t)}  \sum_{m \neq 0,m=-\infty}^{m=\infty} W\left(\frac{m}{2},e\right) \cos(2 \theta- m t ).
$$

\begin{figure}
\centering
\includegraphics[width = 0.8 \textwidth]{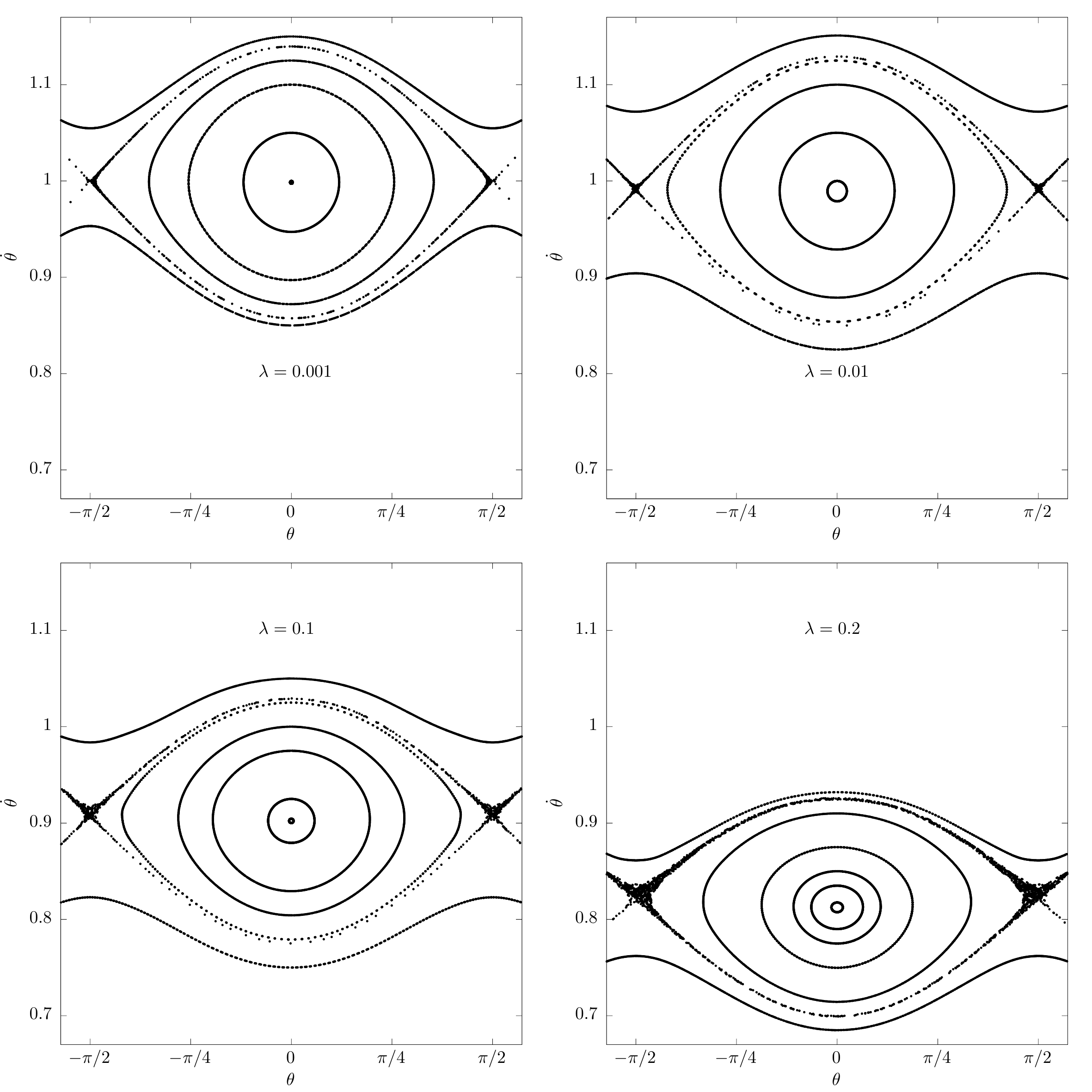}
\caption{The phase space portraits (stroboscopic surfaces of section) of the model in Eq.~(\ref{eq:model1}) for $B-A=0.01$, $\hat{C}=1.0$, $\Omega=1$, $\nu=1$, $e=0.01$ and $\lambda$ as indicated in each panel. The variation of $\lambda$ affects the system in two ways: a) all resonances are shifted downwards in the $\dot{\theta}$ direction, and b) the width of the resonant stochastic layer increases with $\lambda$.}
\label{fig:physicslambda}
\end{figure}

The parameter $\lambda$, in the above equations, controls the degree of variation in the value of the moment of inertia $C(t)$. The effect of $\lambda$ on the dynamics of the system is depicted in Fig.~\ref{fig:physicslambda}. For fixed values of the remaining control parameters, we show the Poincar\'e surfaces of section for different values of $\lambda$. Varying $\lambda$ alters the phase space structure in two ways: a) all periodic solutions (and associated resonances) are vertically displaced (in  $\dot{\theta}$) and b) the width of the stochastic layer around each resonance increases with $\lambda$.

We will now compute the time parametrisation $t(\tau)$ that recasts the system to the Hamiltonian form of Eq.~(\ref{eq:ham}).  In the present case, $F(t)$ is resonant with the orbital motion. Therefore, we set the variation of the moment of inertia to be in 1:1 resonance with the orbital frequency ($\Omega=1$), we choose $\hat{C}=1$ for simplicity and we expand $F(t)$ up to the 4th order in $\lambda$ as:
$$
F(t) = \lambda \sin (t)- \lambda^2 \frac{ \sin (2t)}{2}+ \lambda^3 \frac{ \sin (t)}{4}+ \lambda^3 \frac{ \sin (3t)}{4} - \lambda^4 \frac{ \sin (2t)}{4} - \lambda^4 \frac{ \sin (4 t)}{8} + \Ord(\lambda^5).
$$

Then we follow the steps presented in Sec.~\ref{subsec:theory2} to obtain the series solution. The details are given below.

\subsubsection{Step 1} The function $w(t) = \int F(t)dt$ defined in Eq.~(\ref{eq:wt}) is
$$
w(t) = - \lambda \cos(t)  + \lambda^2 \frac{ \cos(t)^2}{2} -
\lambda^3 \frac{ \cos(t)}{4}  - \lambda^3 \frac{ \cos(3 t)}{12} +
\lambda^4 \frac{ \cos(t)^2}{4} + \lambda^4 \frac{ \cos(4 t)}{ 32} + \Ord(\lambda^5).
$$

\subsubsection{Step 2} The series of its exponential $e^{w(t)}$ up to order 4 in $\lambda$ reads
\begin{align}
e^{w(t)} &= 1 + \frac{\lambda^2}{2} + \frac{13 \lambda^4}{32} - \lambda  \cos(t)  + \lambda^2 \frac{ \cos(2 t)}{2} - \lambda^3 \frac{ 3  \cos(t)}{4} \nonumber\\
  &- \lambda^3 \frac{ \cos(3 t)}{4} + \lambda^4 \frac{ \cos(2 t)}{ 2} + \lambda^4  \frac{\cos(4
  t)}{8} + \Ord(\lambda^5).\nonumber
\end{align}

\subsubsection{Step 3} In order to avoid secular terms in our solution we can fix the constant
$C_1$ so that:
$$
e^{C_1} = \frac{1}{1+ \frac{\lambda^2}{2 } + \frac{13
\lambda^4}{32 }  +\Ord(\lambda^{5})}.
$$
Therefore, we obtain:
$$
e^{w(t)+C_1} = 1 - \lambda  \cos(t)  + \lambda^2  \frac{\cos(2 t)}{ 2 }  -  \lambda^3 \frac{ \cos(t)}{4 } - \lambda^3 \frac{ \cos(3 t)}{4 } + \lambda^4 \frac{ \cos(2 t)}{4 } + \lambda^4  \frac{ \cos(4 t)}{8} +\Ord(\lambda^{5}).
$$

\subsubsection{Step 4}

The time parametrisation $\tau(t)$ is obtained as
$$
\tau = t - \lambda \sin(t) + \lambda^2 \frac{\sin(2 t)}{ 4 }  -
\lambda^3 \frac{ \sin(t)}{4 } - \frac{\lambda^3 \sin(3 t)}{12 } +
\frac{\lambda^4 \sin(2 t)}{8 }  + \frac{ \lambda^4 \sin(4 t)}{32} +\Ord(\lambda^{5}).
$$

\subsubsection{Step 5}

Finally, the inverse function $t(\tau)$ is obtained with a series inversion method; the result up to the 4th order in $\lambda$ is the following:
\begin{equation}
t = \tau + \lambda \sin(\tau) + \lambda^2 \frac{\sin(2 \tau)}{4} +  \lambda^3 \frac{\sin( \tau)}{4 } +  \lambda^3 \frac{\sin(3 \tau)}{12} + \lambda^4 \frac{\sin(2 \tau)}{8} + \lambda^4 \frac{ \sin(4 \tau)}{32} + \Ord(\lambda^{5}).
\label{eq:solutionex1}
\end{equation}

\begin{figure}
\centering
\includegraphics[width = 0.45 \textwidth]{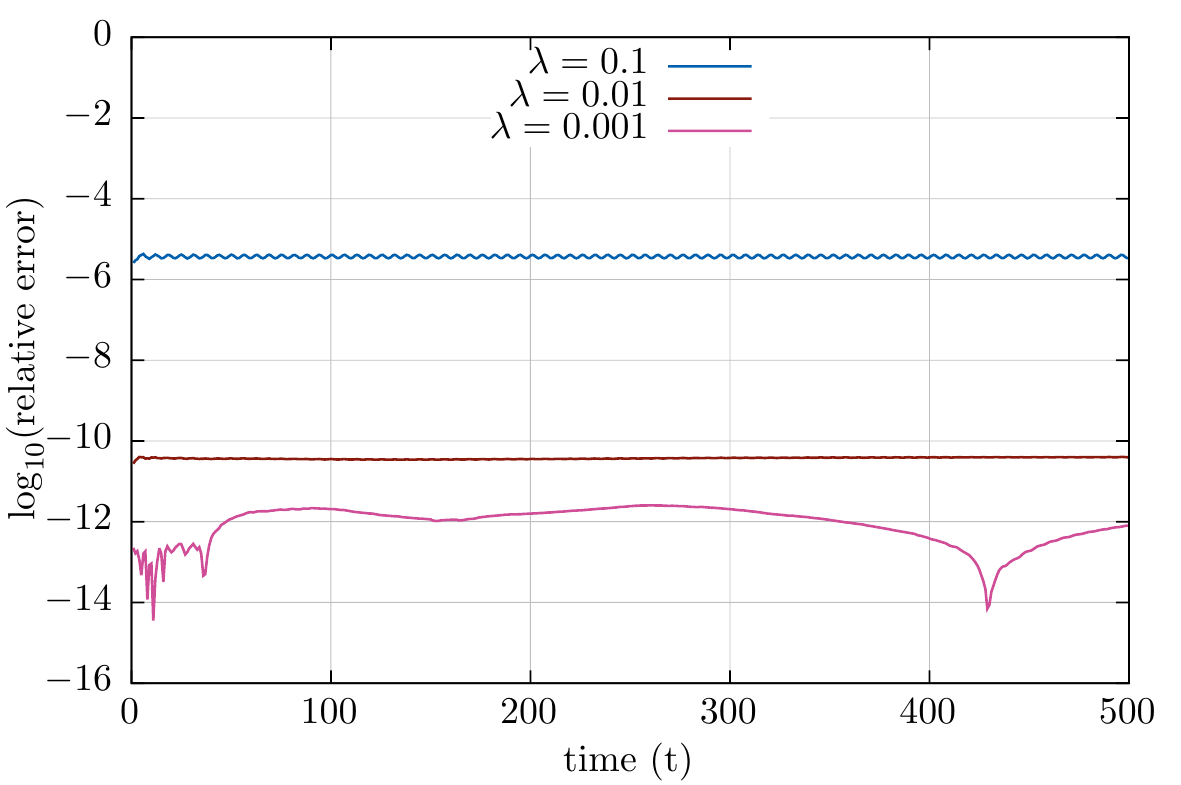}
\includegraphics[width = 0.45 \textwidth]{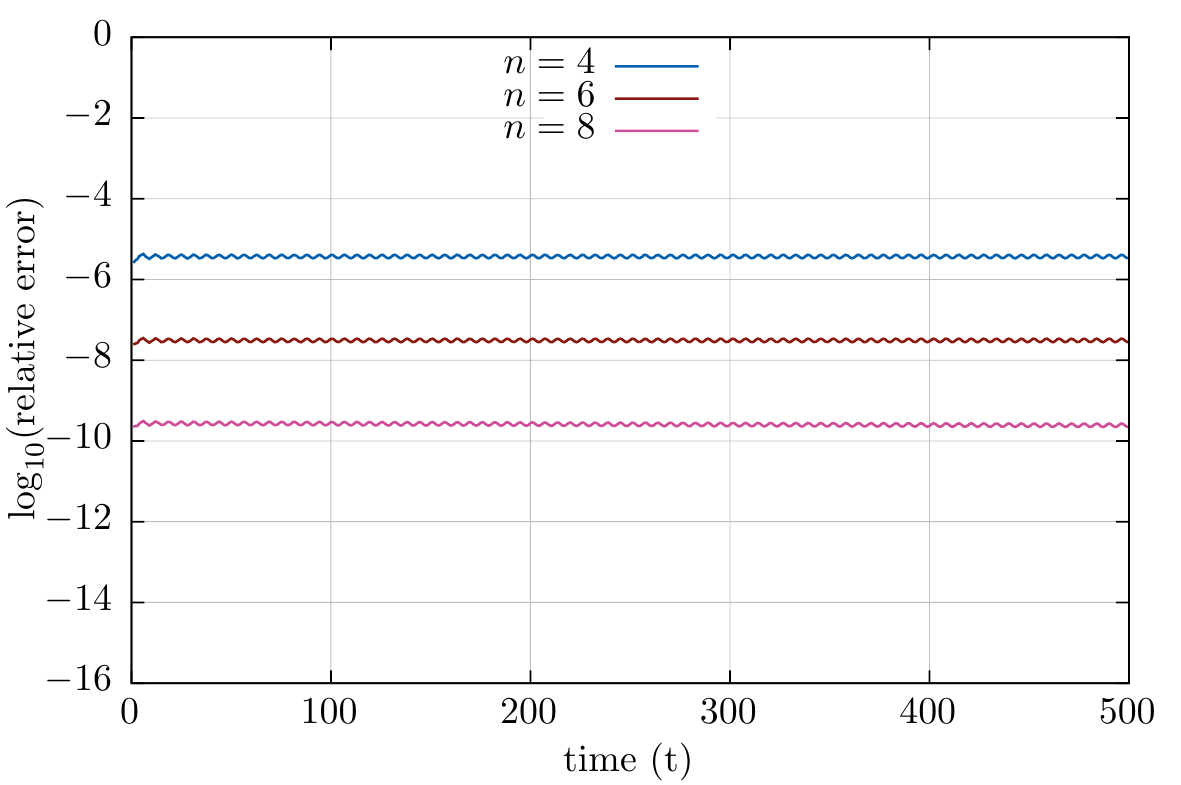}
\caption{The relative error in the time parametrisation between the series solution and the numerical integration: i) for the 4th order solution and different values of $\lambda$ (left) and ii) for different values of the order $n$ and $\lambda=0.1$ (right).}
\label{fig:example1relerror}
\end{figure}

In order to test the analytical series of the time parametrisation we compare it with the numerical integration of Eq.~(\ref{eq:deqtimepar}). In the left panel of Fig.~\ref{fig:example1relerror} we compare the 4th order series solution with the numerical solution (obtained with an accuracy of $10^{-13}$), for three different values of $\lambda$. We observe that our analytical solution up to order 4 is sufficient to model the time parametrisation with an error less than $10^{-10}$ for values of $\lambda<0.01$. Note also that, due to our free of secular terms construction, the error level is almost constant in time. In fact, the fluctuations in the solution $\lambda=0.001$ are due to the fact that the analytical solution has similar accuracy with our numerical method\footnote{For our numerical integrations we used a Bulirsch-Stoer scheme (\cite{bulirsch1966}, \cite{Press1982}) with an accuracy of $10^{-13}$.}. In the right panel of Fig.~\ref{fig:example1relerror} we compare our series solution, truncated to different orders $n$, with the numerical solution for $\lambda = 0.1$. We observe again the bounded nature of the error, due to the absence of secular terms. We note also that our series solution gains about one order of magnitude in the error for every extra truncation order.

We will use now the time parametrisation to write the system in Hamiltonian form as (compare with \equ{eq:ham}):
\begin{equation}
H(I,\theta,\tau) = \frac{I^2}{2} + V(\theta,t(\tau))
\left(\frac{dt}{d\tau}\right)^2 , \label{eq:hamex1}
\end{equation}
where $I = \dot{\theta} \frac{dt}{d\tau}$ (see Eq.~(\ref{eq:neweq1})) and
\begin{equation}
V(\theta,\tau) = - \frac{3}{4} \frac{\nu (B - A)}{1 + \lambda \cos
(t(\tau))}  \sum_{m \neq 0,m=-\infty}^{m=\infty}
W\left(\frac{m}{2},e\right) \cos(2 \theta- m t(\tau) ).
\label{eq:potentialex1}
\end{equation}

\begin{figure}
\centering
\includegraphics[width = 0.45 \textwidth]{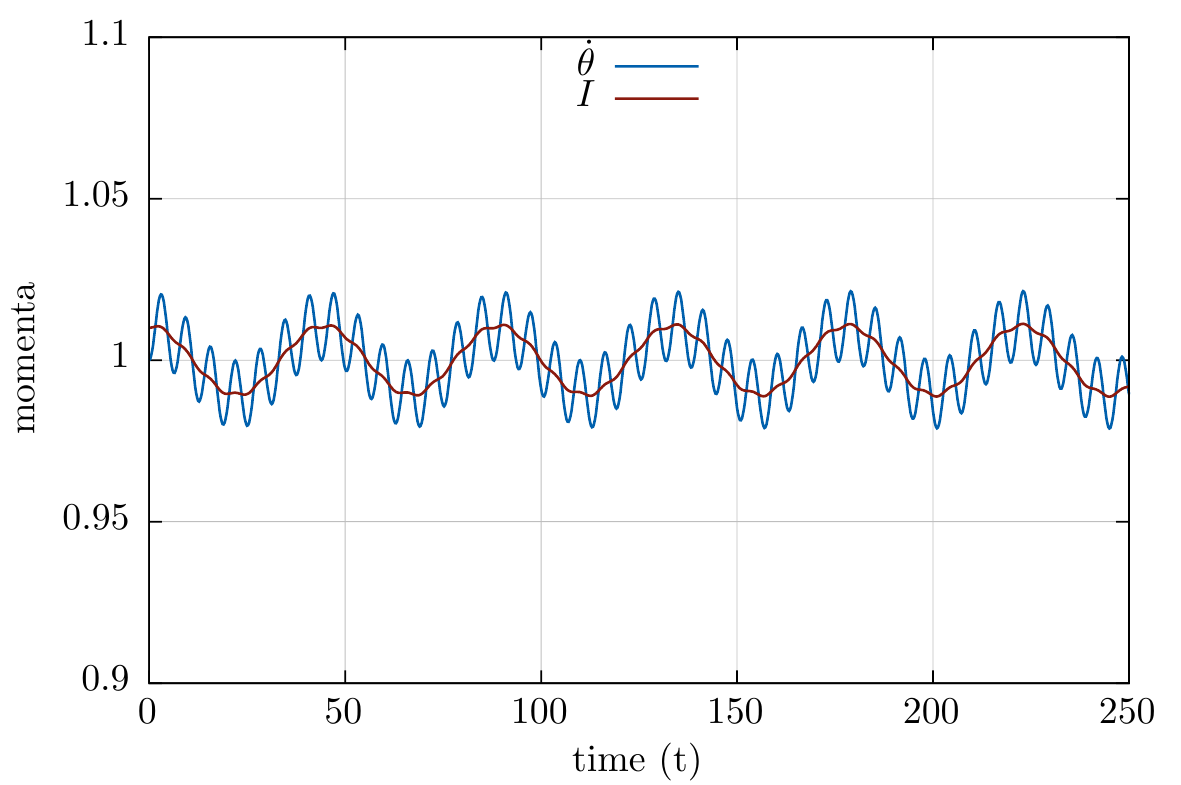}
\includegraphics[width = 0.45 \textwidth]{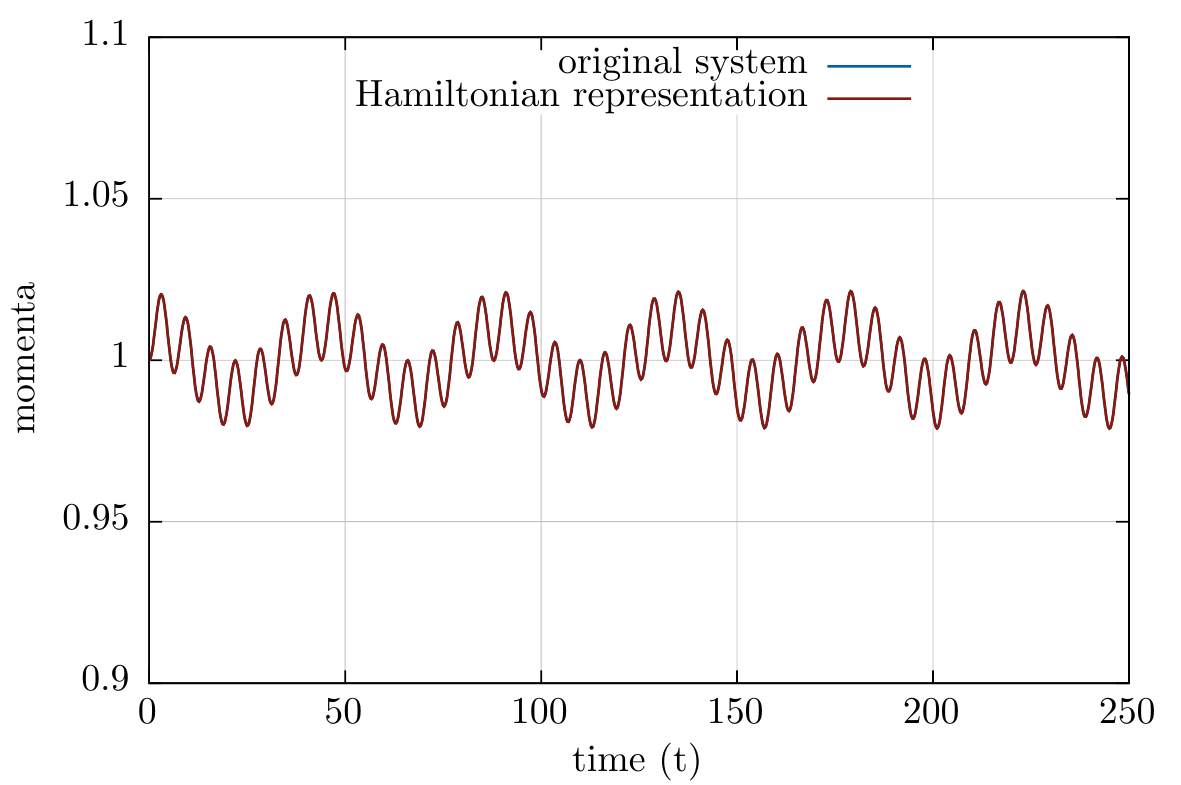}
\caption{Time evolution of the momenta $\dot{\theta}$ and $I$ in the full system and the Hamiltonian representation (left). Taking into account the time parametrisation $t(\tau)$ the two solutions coincide (right).}
\label{fig:example1single}
\end{figure}

In Fig.~\ref{fig:example1single}, left panel, we present the time evolution of $\dot{\theta}$ in the full system and the evolution of $I$ in the Hamiltonian formulation. We note that the action variable $I$ accounts, on the average, for the time variations of $\dot{\theta}$. However, $I$ exhibits a behaviour averaged with respect to short fluctuations, which evolve with the short period of the change of moment of inertia of the satellite. However, the exact behaviour is recovered if we take into account also the time parametrisation $t(\tau)$ of Eq.~(\ref{eq:solutionex1}) (see Fig.~\ref{fig:example1single}, right panel).

\begin{figure}
\centering
\includegraphics[width = 0.7 \textwidth]{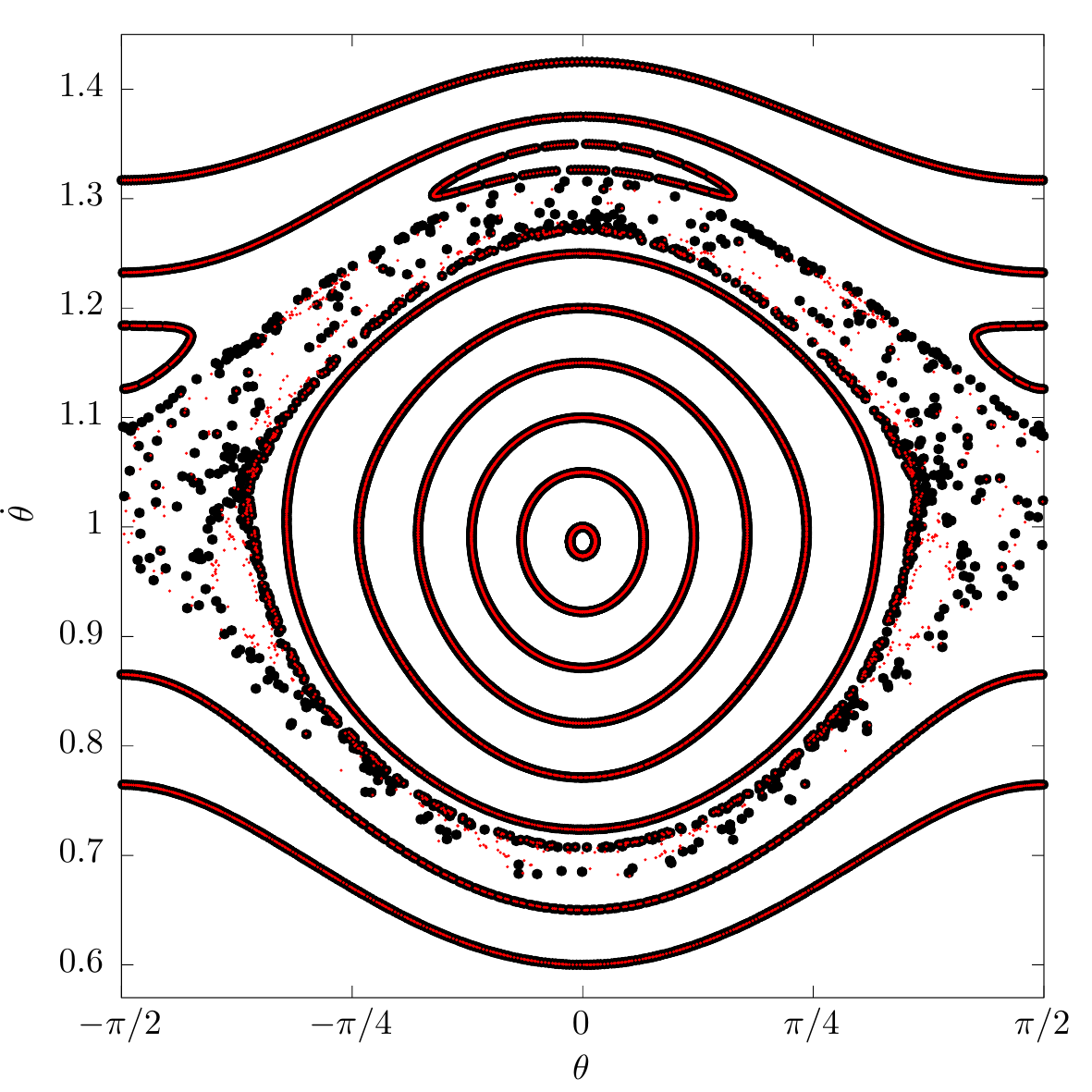}
\caption{The comparison between the Poincar\'e surface of section of the original (non-Hamiltonian) system (black dots) and the surface of section of the Hamiltonian representation transformed in the original variable (red dots).}
\label{fig:example1pss}
\end{figure}

Furthermore, expanding the Hamiltonian (\ref{eq:hamex1}), (\ref{eq:potentialex1}) up to 4th order in both the eccentricity and $\lambda$, we obtain
$$
H(I,I_{\tau},\theta,\tau) =
\frac{I^2}{2} + I_{\tau} +  \sum_{k_0,k_1,k_2,k_3}
c_{k_0k_1k_2k_3} e^{k_0} \lambda^{k_1} {\rm e}^{i (k_3 \theta +
k_4 \tau)}\ , k_0,k_1,k_2,k_3 \in \mathbb N\ , k_0,k_1 \leq 4,
$$
where $I_{\tau}$ is a dummy action associated to the time parametrisation $t(\tau)$. We note that, as customary in Celestial Mechanics, here we introduce one dummy action variable for each independent angle associated with one independent external frequency modulating the system. This implies that the system is formally equivalent to one in which each action-angle pair counts as one more degree of freedom. In our specific example, since $\Omega$ and $\nu$ are in 1:1 resonance, the frequency associated with the dummy action $I_{\tau}$ is also equal to unity. Thus, we end up with a two degrees of freedom Hamiltonian system, whose phase space can be studied by means of a Poincar\'e surface of section. In Fig.~\ref{fig:example1pss} the results are compared with those obtained from the numerical integration of the original (non-Hamiltonian) system, i.e. a stroboscopic map computed by Eq.~(\ref{eq:euler}). To create the stroboscopic map we record the values of $\theta$ and $\dot{\theta}$ over fixed time intervals. Assuming a general resonant case, where $\Omega / \nu = p / q$ (with $p \in\integer$, $q\in\integer\backslash\{0\}$), the stroboscopic map is defined every $T = 2 q \pi / \nu$.

In Fig. \ref{fig:example1pss} we present the stroboscopic map for the 1:1 resonant case. Despite its formal appearance, once studied with the stroboscopic map, the original system retains essentially a Hamiltonian character, i.e. it exhibits typical features (regular KAM curves, islands of stability, resonances, chaotic layers, etc.) of a genuinely Hamiltonian system. Specifically, the Poincar\'e surface of section, computed using the Hamiltonian system, matches very well the section of the original system, if the variable $I$ is back-transformed to $\dot{\theta}$. Discrepancies appear only in the chaotic domain, where the same initial conditions give different evolutions in the original and the time parametrized, and the truncated Hamiltonian system. This is only due to the exponential growth of the error in the computation of the chaotic trajectories.

\subsection{Example of $C(t)$ periodic, non-resonant with the orbit.}

In this section, a non-resonant case will be treated, in which we choose a non-commensurable ratio of orbital to moment of inertia variation frequencies by setting $\Omega=\sqrt{2}$ and we expand $F(t)$ up to fourth order in $\lambda$:

\begin{align}
F(t) &= \lambda  \sin (\sqrt{2} t)- \lambda^2 \frac{ \sin (2 \sqrt{2} t)}{2}+ \lambda^3 \frac{ \sin (\sqrt{2} t)}{4}+ \lambda^3 \frac{ \sin (3 \sqrt{2} t)}{4} - \lambda^4 \frac{ \sin (2 \sqrt{2} t)}{4} \nonumber \\
&- \lambda^4 \frac{ \sin (4 \sqrt{2} t)}{8}  + \Ord(\lambda^{5}).
\end{align}
Following the discussion of the previous section, the time parametrisation in this case reads:
\begin{align}
t &= \tau + \lambda \frac{ \sin(\sqrt{2} \tau)}{\sqrt{2}} + \lambda^2 \frac{\sin(2 \sqrt{2} \tau)}{4 \sqrt{2}} +
 \lambda^3 \left( \frac{\sin(\sqrt{2} \tau)}{4 \sqrt{2}} + \frac{\sin(3 \sqrt{2} \tau)}{12 \sqrt{2}} \right) \nonumber \\
 &+  \lambda^4 \left( \frac{\sin(2 \sqrt{2} \tau)}{8 \sqrt{2}} + \frac{ \sin(4 \sqrt{2} \tau)}{32 \sqrt{2}} \right)  + \Ord(\lambda^{5}).\nonumber
\end{align}

\begin{figure}
\centering
\includegraphics[width = \textwidth]{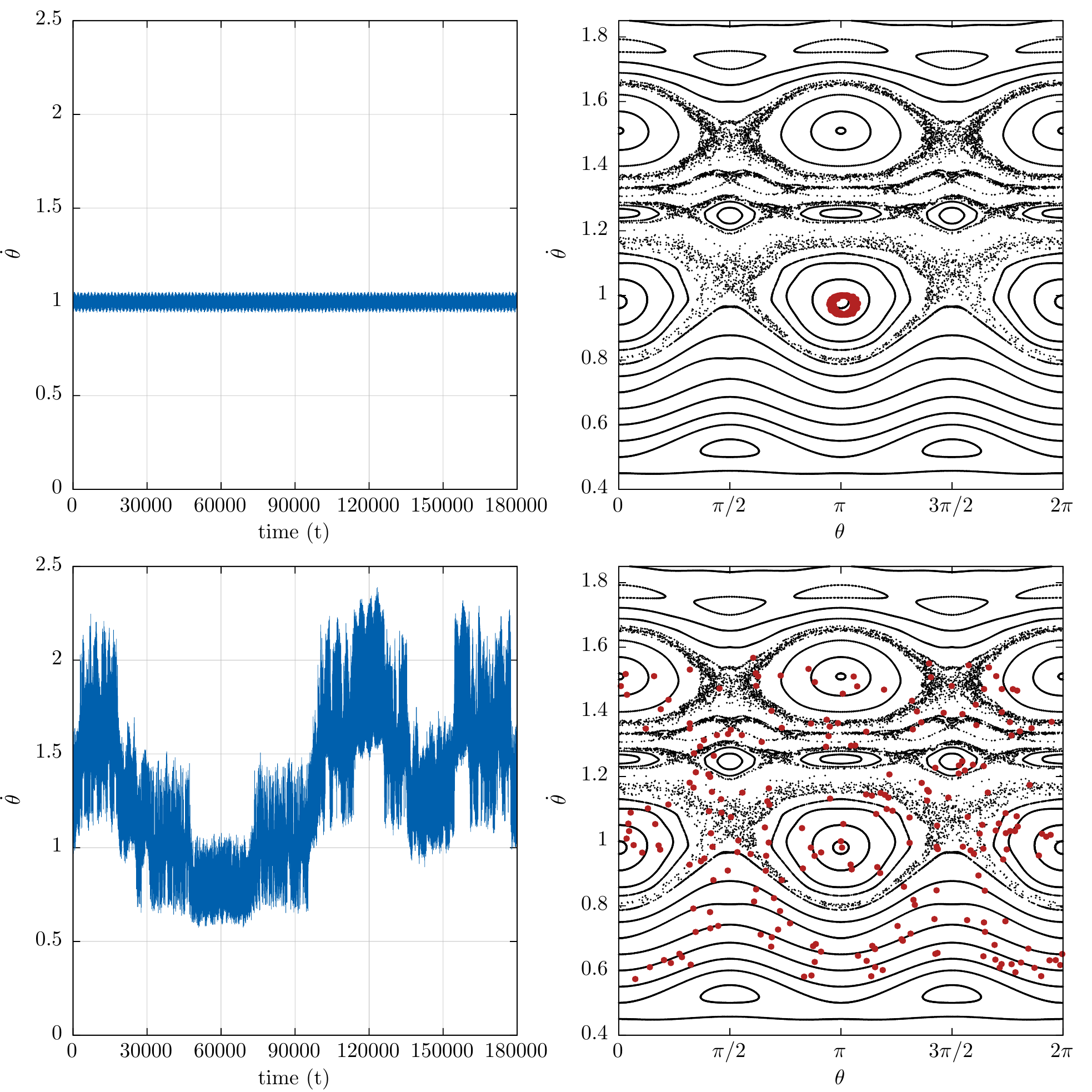}
\caption{In the left panels we present the time evolution of the
momentum $\dot{\theta}$ in the non-resonant case $\Omega = \sqrt{2}$ for the samples $\lambda = 0.01$ (top) and $\lambda = 0.2$ (bottom). In the right panels we present a double section of the full model (red dots) superposed to the usual Poincar\'e surface of section of the conservative model (black dots).}
\label{fig:example2long}
\end{figure}
Expanding the Hamiltonian up to 4th order in both the eccentricity and $\lambda$ we obtain the following Hamiltonian:
\begin{eqnarray}
\label{eq:isochronous}
H(I,I_{\tau},I_{\tau'},\theta,\tau,\tau') = \frac{I^2}{2} + I_{\tau} + \sqrt{2} I_{\tau'} +  \sum_{k_0,k_1,k_2,k_3,k_4} c_{k_0k_1k_2k_3k_4} e^{k_0} \lambda^{k_1} {\rm e}^{i (k_2 \theta + k_3 \tau + k_4 \tau')}\,  \\
 \quad k_0,k_1,k_2,k_3,k_4 \in \mathbb N\ \nonumber ,
\end{eqnarray}
where $\tau'=\sqrt{2} \tau$ and $I_{\tau'}$ is a dummy action associated to $\tau'$. Note that in this case, where the orbital and the moment of inertia variation frequencies are not commensurable, two independent dummy actions have to be introduced in the Hamiltonian description of the system. Therefore, an extra degree of freedom is introduced and Eq.~(\ref{eq:isochronous}) represents an isochronous system in which diffusion could appear (\cite{gallavotti1998}) through the extra degree of freedom.

In Fig.~\ref{fig:example2long} we present the evolution of the system in two cases: for $\lambda=0.01$ (top panels) and for $\lambda=0.2$ (bottom panels). In the left panels, we present the time evolution of the angular velocity $\dot\theta$ by integrating the equation of motion of the original system. In the case $\lambda=0.01$ the momentum seems to be bounded, while in the case $\lambda=0.2$ it appears to move more erratically, getting temporarily trapped into particular regions of the phase space. In order to qualitatively describe this feature, we produce a double-section of the system. In this case, similarly to the stroboscopic map, we register the values of ($\theta,\dot\theta$) when
\begin{equation}
mod(\tau,2 \pi) \approx  mod(\tau', 2 \pi) \approx 0.
\end{equation}
For comparison, in both cases, we superpose the double section to the usual stroboscopic map in the case of $\lambda=0$. We observe that for $\lambda=0.01$ the double section is similar to an invariant curve of the original system. On the other hand, in the case $\lambda=0.2$ the system seems to wander in the phase space in a chaotic manner.

\subsection{Example of $C(t)$ periodic with tidal dissipative force}\label{sec:tidal}

In this example we study the case where $C(t)$ is periodic and in the system acts also a tidal torque of the form
$$
N_{z(tidal)}(\dot{\theta}) = \mu + a \dot{\theta},
$$
for some real constants $\mu$, $a$. We focus on the case $\mu>0$, $a<0$, motivated by models of tidal interactions as e.g in \cite{goldreich1966}, \cite{efroimsky2009}, \cite{noyelles2014}.

In this case the equation of motion takes the form

\begin{equation}
\ddot{\theta} =  - \frac{3}{2} \frac{\nu (B - A)}{\hat{C} + \lambda \cos (\Omega t)}  \sum_{m \neq 0,m=-\infty}^{m=\infty} W\left(\frac{m}{2},e\right) \sin(2 \theta- m t ) + \frac{\lambda \Omega \sin{(\Omega t)}}{\hat{C} + \lambda \cos(\Omega t)}\dot{\theta} + \mu + a \dot{\theta},
\label{eq:example3fullsystem}
\end{equation}

and it is equivalent to Eq.~(\ref{eq:originalsystem}) with
\begin{align}
G(\theta,t) &=  - \frac{3}{2} \frac{\nu (B - A)}{\hat{C} + \lambda \cos (\Omega t)}  \sum_{m \neq 0,m=-\infty}^{m=\infty} W\left(\frac{m}{2},e\right) \sin(2 \theta- m t ) + \mu,\nonumber\\
F(t) &=  a + \frac{\lambda \Omega \sin{(\Omega t)}}{\hat{C} +
\lambda \cos(\Omega t)}.\nonumber
\end{align}

The potential function $V(\theta,t)$ associated with $G(t)$ is
\begin{equation}
V(\theta,t) = - \frac{3}{4} \frac{\nu (B - A)}{\hat{C} + \lambda
\cos (\Omega t)}  \sum_{m \neq 0,m=-\infty}^{m=\infty}
W\left(\frac{m}{2},e\right) \cos(2 \theta- m t ) - \mu \theta.
\label{eq:potentialV}
\end{equation}
Here, we apply the generalized version of our method, developed in Sec. \ref{subsec:fulltheory}, in order to calculate the time parametrisation and we consider a resonant case with $\Omega=1$.

\subsubsection{Steps 1-3}
The first three steps of the procedure are identical to the resonant derivation, but we will keep terms only up to second order in $\lambda$ and choose $\hat{C}=1$ for simplicity:
$$
e^{w(t)+C_1} = 1 - \lambda \cos(t)  + \lambda^2  \frac{\cos(2 t)}{2}  + \Ord(\lambda^{3}).
$$

\subsubsection{Step 4}
In the next step we compute the time integral as follows:
\begin{align}
\tau + C_2 &= \int e^{at + w(t) + C_1} dt = \int e^{at} \left( 1 - \lambda \cos(t)  + \lambda^2  \frac{\cos(2 t)}{2} \right)  + \Ord(\lambda^{3})= \nonumber \\
&= \frac{e^{a}}{a} \left( 1 - \frac{4 a^2 \lambda \cos{t}}{C_A}
-\frac{a^4 \lambda \cos{t}}{C_A} + \frac{ a^2 \lambda^2 \cos{2
t}}{2 C_A} + \frac{ a^4 \lambda^2 \cos{2 t}}{2 C_A}  \right. \\
\nonumber & \left.  - \frac{4 a \lambda \sin{t}}{C_A} - \frac{a^3
\lambda \sin{t}}{C_A} +  \frac{a \lambda^2 \sin{2 t}}{C_A} +
\frac{a^3 \lambda^2 \sin{2 t}}{C_A} \right) + \Ord(\lambda^{3}),\nonumber
\end{align}
where $C_A=4 + 5 a^2 + a^4$.

\subsubsection{Step 5}
In the last step of the procedure we invert the series $\tau(t)$ to obtain $t(\tau$). Up to second order in $\lambda$, we obtain:
\begin{align}
t &= L + \Pi + \frac{4 a \lambda \cos{L}}{C_A} + \frac{a^3 \lambda \cos{L}}{C_A} - \frac{16 a \lambda^2 \cos{L}}{C_A^2 C_B} - \frac{8 a^3 \lambda^2 \cos{L}}{C_A^2 C_B} - \frac{a^5 \lambda^2 \cos{L}}{C_A^2 C_B} \nonumber \\
& + \frac{16 a \lambda^2 \cos^2{L}}{C_A^2} + \frac{8 a^3 \lambda^2 \cos^2{L}}{C_A^2} + \frac{ a^5 \lambda^2 \cos^2{L}}{C_A^2} - \frac{4 a \lambda^2 \cos{2 L}}{C_A^2} + \frac{ 2 a^3 \lambda^2 \cos{2 L}}{C_A^2} \nonumber\\
& + \frac{7 a^5 \lambda^2 \cos{2 L}}{4 C_A^2} + \frac{a^7 \lambda^2 \cos{2 L}}{4 C_A^2} - \frac{a \lambda^2 \cos{2 L}}{2 C_A} - \frac{a^3 \lambda^2 \cos{2 L}}{2 C_A} + \frac{4 \lambda \sin{L}}{C_A} \nonumber\\
& + \frac{a^2 \lambda \sin{L}}{C_A} + \frac{16 a^2 \lambda^2 \sin{L}}{C_A^2 C_B} + \frac{8 a^4 \lambda^2 \sin{L}}{C_A^2 C_B} + \frac{ a^6 \lambda^2 \sin{L}}{C_A^2 C_B} + \frac{16 \lambda^2 \cos{L} \sin{L}}{C_A^2} \\
&  - \frac{ 8 a^2 \lambda^2 \cos{L} \sin{L}}{C_A^2} - \frac{7 a^4 \lambda^2 \cos{L} \sin{L}}{C_A^2} - \frac{a^6 \lambda^2 \cos{L} \sin{L}}{C_A^2} - \frac{16 a \lambda^2 \sin^2{L}}{C_A^2} \nonumber \\
& - \frac{8 a^3 \lambda^2 \sin^2{L}}{C_A^2}  - \frac{a^5 \lambda^2 \sin^2{L}}{C_A^2} + \frac{8 a^2 \lambda^2 \sin{2 L}}{C_A^2} + \frac{4 a^4 \lambda^2 \sin{2 L}}{C_A^2} + \frac{a^6 \lambda^2 \sin{2 L}}{2 C_A^2} \nonumber \\
& - \frac{\lambda^2 \sin{2 L}}{C_A} - \frac{a^2 \lambda^2 \sin{2
L}}{C_A}  + \Ord(\lambda^{3}), \label{eq:step53}
\end{align}
where $C_A=4 + 5 a^2 + a^4$, $C_B = 1 + a \tau$ and $L=\log (1+a \tau)/ a$. In addition $\Pi$ is given by:
$$
\Pi(a,\tau) = \frac{a \lambda (-4 (4 + a^2) C_A C_B + (4 + a^2)^2 (C_B^2 + a^2 (-2 + C_B^2)) \lambda + 2 (1 + a^2) C_A C_B \lambda)}{4 C_A^2 C_B^2}.
$$

\begin{figure}
\centering
\includegraphics[width = 0.45 \textwidth]{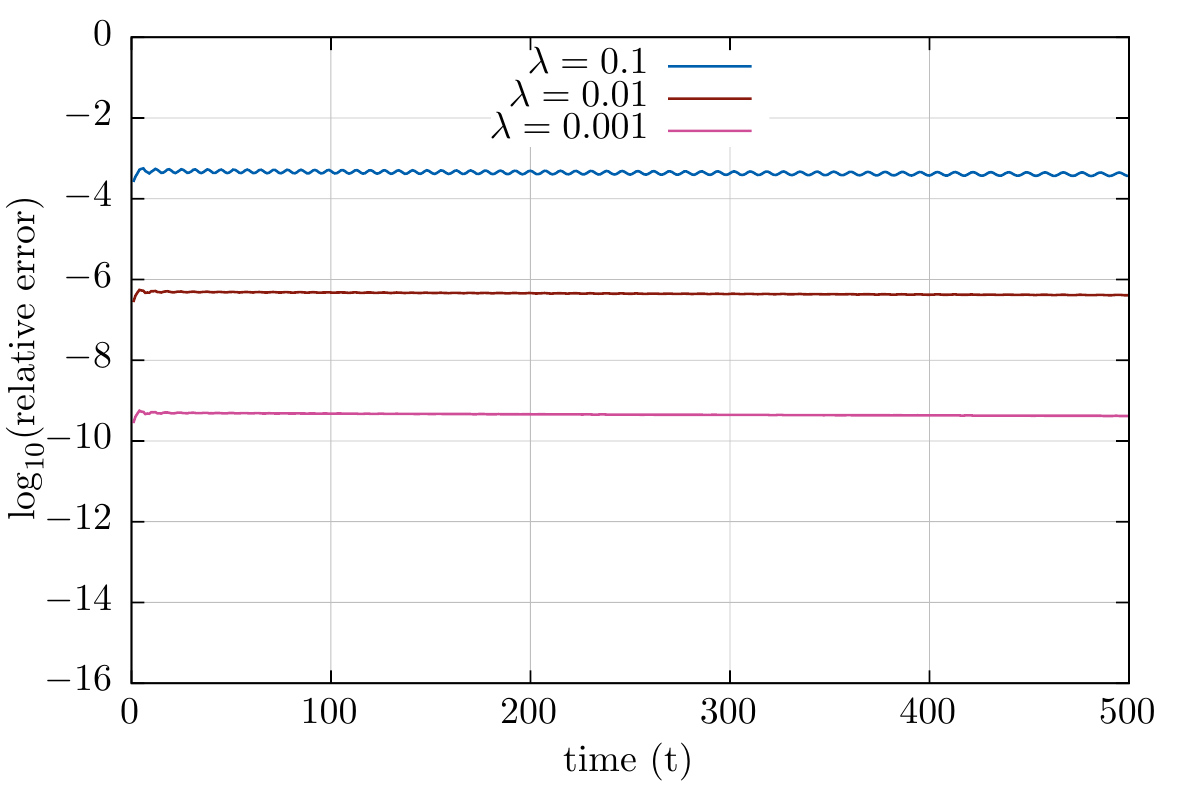}
\includegraphics[width = 0.45 \textwidth]{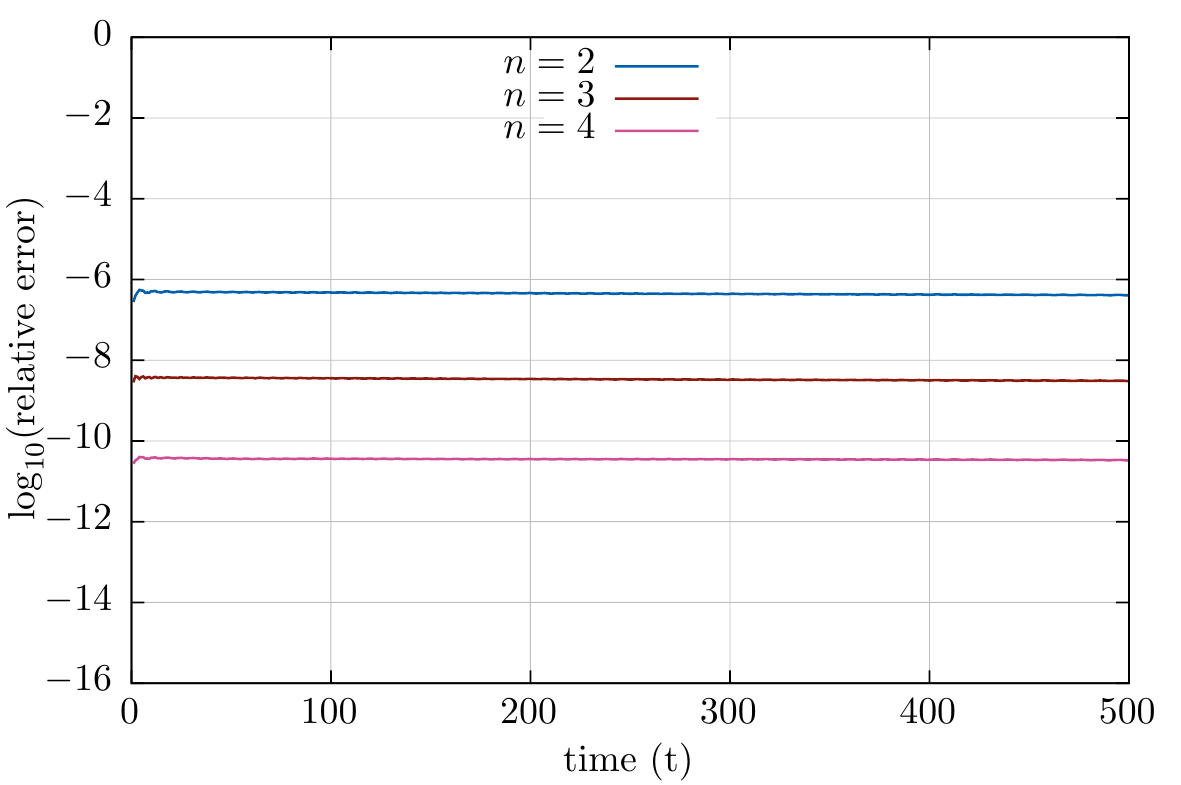}
\caption{The relative error in the time parametrisation between the series solution and the numerical integration: i) for the second order solution and different values of $\lambda$ (left) and ii) for different values of order $n$ and $\lambda=0.01$ (right). In all computations $a=-10^{-3}$.}
\label{fig:example3relerror}
\end{figure}

Similar to the other cases, the first test we perform is the accuracy of the analytical series for the time parametrisation. In the left panel of Fig. \ref{fig:example3relerror} we compare the second order analytical series with the numerical solution for the time parametrisation equation (Eq. (\ref{eq:deqtimepar})) for a fixed value of $a=-10^{-3}$ and different values of $\lambda$. It is evident that the behaviour of the analytical series solution is pretty satisfactory and similar to the solution of the resonant case. Furthermore, in the right panel of Fig.~\ref{fig:example3relerror} we compare the analytical and numerical solutions for a fixed value of $\lambda = 0.01$ and different truncation orders $n$. Again, the results are in excellent agreement and the general behaviour matches with that of the resonant case.

The analytical calculation of the parametrisation allows us to write Eq.~(\ref{eq:example3fullsystem}) in a Hamiltonian form as
\begin{equation}
H = \frac{I^2}{2} + \left(\frac{dt}{d\tau}\right)^2
V(\theta,t(\tau)). \label{acca}
\end{equation}

\begin{figure}
\centering
\includegraphics[width = 0.38 \textwidth]{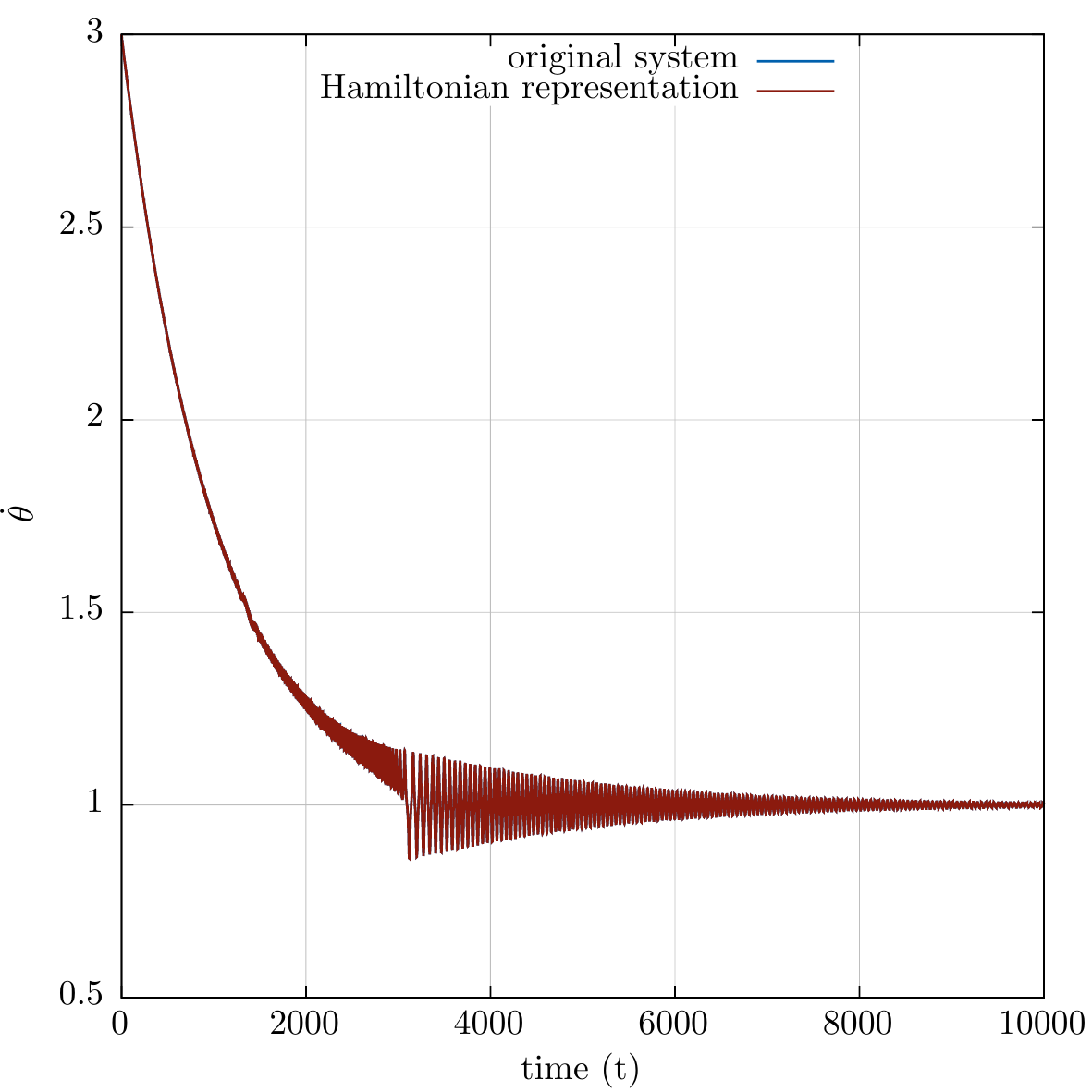}
\includegraphics[width = 0.45 \textwidth]{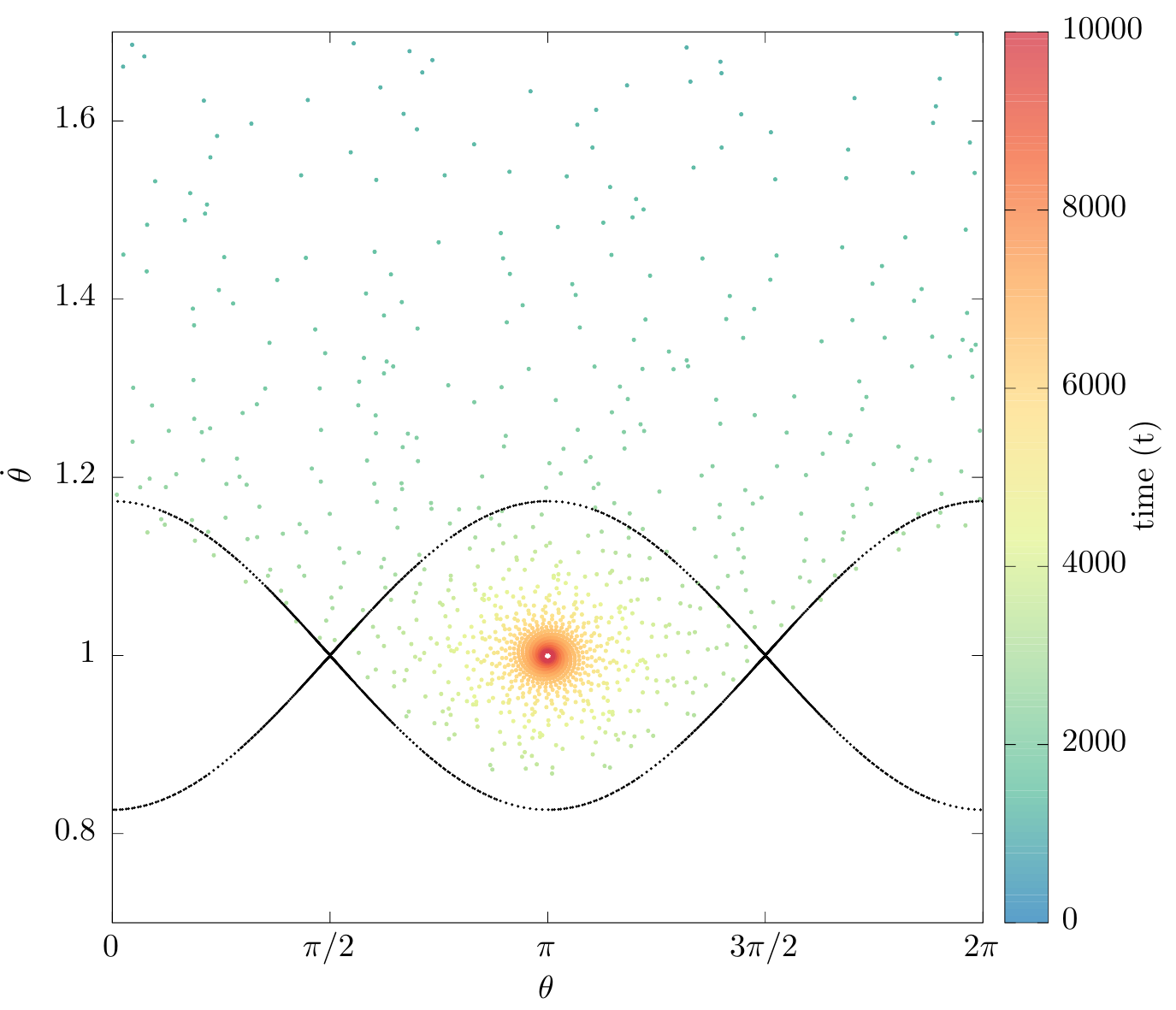}
\caption{Time evolution of the momenta $\dot{\theta}$ and
$I/(dt/d\tau)$ in the original system
(Eq.~(\ref{eq:example3fullsystem})) and the Hamiltonian
representation (see Eq.s \equ{acca}, (\ref{eq:step53}),
(\ref{eq:potentialV})) in the case of a capture into the 1:1
resonance (left). The stroboscopic Poincar\'e map for this scenario that shows the capture into the synchronous resonance. Points are coloured according to their time (t) label (right).}
\label{fig:example3single}
\end{figure}

As a final application, we use our technique to describe the evolution of the system in the case $a=-10^{-3}$, $\mu= 10^{-3}$ and $\lambda = 10^{-4}$. In the left panel of Fig.~\ref{fig:example3single} we present the time evolution of the angular momentum, assuming also that the body orbits on an ellipse with eccentricity $e=0.01$. The blue curve (which in fact overlaps and it is hidden  by the red curve) corresponds to the direct numerical integration of the original, non-conservative, system described by Eq.~(\ref{eq:example3fullsystem}). The red curve is obtained from the numerical integration of the Hamiltonian system given by Eq.~\equ{acca} with $t(\tau)$ as in Eq.~(\ref{eq:step53}) and $V$ in Eq.~(\ref{eq:potentialV}), after applying the back transformation computed by the time parametrisation. This numerical example demonstrates  that our technique allows for an accurate Hamiltonian description of the capture of the system into resonance.

The capture into the synchronous resonance is also depicted in the right panel of Fig.~\ref{fig:example3single}, with the help of a stroboscopic Poincar\'e map. Due to form of the dissipative forces, all phase space structures of the conservative model now disappear. However, by using a colour scale to draw the points in the $(\theta,\dot{\theta})$ plane according to the different phases of evolution (time windows), the capture process becomes evident. In particular, the capture within the separatrix of the 1:1 resonance of the unperturbed system ($\lambda = \mu = a =0$) takes place at a time around $t=3500$, which is in agreement with the evolution of the $\dot{\theta}$ shown in the first panel.

\section{Conclusions}\label{sec:conclusions}

In this work we considered a model problem which describes the rotational motion of a non-rigid body under non-conservative effects. In particular, we have studied the flexibility of the body, which is modelled through the time variation of the moments of inertia, and the effect of the tidal torque raised by a central body. In the considered examples, the equations of motion involve an angular velocity-dependent term that destroys the symplectic structure. However, such structure can be conveniently recovered with the aid of a suitable time parametrisation, which can be computed in series form. In Sec.~\ref{sec:theory} we provided the methodology to obtain this parametrisation in various cases. Furthermore, we tested our theory in three different samples. Although we lack a formal proof, the results in all three examples indicate that the method converges for all practical purposes and can be applied in real case scenarios. The main advantage of casting this kind of systems into a Hamiltonian form is that we open a window of applying classical knowledge of conservative systems in applications to dissipative cases. Although the model considered in this work is a special one, the described methods can be possibly extended to more general situations. Time varying non-conservative forces appear in many different contexts of physics and astronomy. Regarding, in particular, applications in astrodynamics, the methods presented above lend themselves quite conveniently in the study of the tidal dynamics of natural satellites with dissipation (e.g. with liquid cores or other types of mechanical friction), but also of artificial satellites, as for example, the motion under a time varying atmospheric drag, the variable mass problem (due, e.g., to fuel consumption), or even oscillating solar wind and/or solar radiation pressure effects.

\vglue1cm

{\bf Acknowledgements.} A.C. and G.P. were partially supported by GNFM-INdAM and by the Stardust Marie Curie Initial Training Network, FP7-PEOPLE-2012-ITN, Grant Agreement 317185. I.G. was supported by the Stardust Marie Curie Initial Training Network, FP7-PEOPLE-2012-ITN, Grant Agreement 317185.

\end{document}